\documentclass[%
 reprint,
 amsmath,amssymb,
 aps,
 pra,
]{revtex4-2}

\usepackage{graphicx}
\graphicspath{{figs/}}
\usepackage{dcolumn}
\usepackage{bm}

\begin{document}

\preprint{APS/123-QED}

\title{Locally Suppressed Transverse-Field Protocol for Diabatic Quantum Annealing}

\author{Louis Fry-Bouriaux}
\email{l.fry-bouriaux@ucl.ac.uk}
\affiliation{%
 London Centre for Nanotechnology, University College London, WC1H 0AH London, UK
}%
\author{Daniel T. O'Connor}
\affiliation{%
 London Centre for Nanotechnology, University College London, WC1H 0AH London, UK
}%
\author{Natasha Feinstein}
\affiliation{%
 London Centre for Nanotechnology, University College London, WC1H 0AH London, UK
}%
\author{Paul A. Warburton}
\affiliation{%
 London Centre for Nanotechnology, University College London, WC1H 0AH London, UK
}%
\affiliation{%
 Department of Electronic \& Electrical Engineering, University College London, WC1E 7JE London, UK
}%

\date{\today}

\begin{abstract}
Diabatic quantum annealing (DQA) is an alternative algorithm to adiabatic quantum annealing (AQA) that can be used to circumvent the exponential slowdown caused by small minima in the annealing energy spectrum. We present the locally suppressed transverse-field (LSTF) protocol, a heuristic method for making stoquastic optimization problems compatible with DQA. We show that, provided an optimization problem intrinsically has magnetic frustration due to inhomogeneous local fields, a target qubit in the problem can always be manipulated to create a double minimum in the energy gap between the ground and first excited states during the evolution of the algorithm. Such a double energy minimum can be exploited to induce diabatic transitions to the first excited state and back to the ground state. In addition to its relevance to classical and quantum algorithmic speed-ups, the LSTF protocol enables DQA proof-of-principle and physics experiments to be performed on existing hardware, provided independent controls exist for the transverse qubit magnetization fields. We discuss the implications on the coherence requirements of the quantum annealing hardware when using the LSTF protocol, considering specifically the cases of relaxation and dephasing. We show that the relaxation rate of a large system can be made to depend only on the target qubit presenting new opportunities for the characterization of the decohering environment in a quantum annealing processor.
\end{abstract}

\maketitle


\section{Introduction}

Quantum annealing is widely regarded as a promising technique for solving difficult optimization problems\cite{Finnila1994,Kadowaki1998}. Since its inception, adiabatic quantum annealing (AQA) is anticipated to yield a computational speedup via quantum tunneling processes, whereby the physical algorithm can overcome tall and narrow energy barriers\cite{Lanting2010, Johnson2011}. Although much progress has been made in understanding and characterizing the role that quantum tunneling plays in AQA\cite{Boixo2015, Denchev2016}, it remains a hotly contested topic \cite{Rieffel2014, Albash2015, Muthukrishnan2016, Albash2018}. Particularly with regard to anticipated computational speed-ups, only specific instances are known where an advantage is observed using AQA algorithms\cite{Denchev2016, Albash2018, King2021} while many classical algorithms, particularly quantum Monte Carlo (QMC) methods, outperform AQA\cite{Denchev2016, Andriyash2017}.

The main bottleneck encountered in AQA is the occurrence of small local minima in the energy spectrum at some point during an anneal. To avoid a transition to an excited state via a local minimum, the annealing schedule can only be run as fast as allowed by the adiabatic theorem, which can be expressed approximately, in the case of traditional transverse-field AQA, as $t_f \propto \Delta^{-2}$ where $t_f$ is the anneal run time, and $\Delta$ is the global minimum of the energy gap. This inverse squared scaling of the anneal run time with the size of the minimum gap severely impacts the performance of AQA when applied to so-called small-gap problems, which thereby have a reputation for being `difficult to solve' in this context\cite{Albash2018}. This limitation is at odds with the statement that AQA can benefit from quantum tunneling in general, since tall and wide energetic barriers necessarily lead to the formation of a very small energy gap, and hence a high error rate for the AQA algorithm in those cases. Many studies using exact and approximate techniques for systems in the thermodynamic limit have argued that such tall and wide energetic barriers are a common feature of annealing problems involving a first-order transition\cite{Jorg2008, Jorg2009, Jorg2010, Altshuler2010, Knysh2016}.

An alternative to AQA that is attracting much attention recently involves the use of local minima in the energy spectrum to diabatically induce transitions between states. In perhaps the most promising technique currently, diabatic quantum annealing (DQA), a diabatic cascade of Landau-Zener (LZ) transitions\cite{Muthukrishnan2016} is utilized such that the system can be initialized in the ground state and finalized in the ground state via a shortcut through the excited state spectrum, leading to prospects of quantum-enabled computational speedups\cite{BERRY1995, Munoz-Bauza2019}. Although this technique poses new experimental challenges and hardware requirements, there are notable benefits over AQA besides circumventing the small-gap bottleneck, the most attractive being universality and relative simplicity compared to gate-model implementations\cite{Crosson2020}. There are already known problems that are suitable for large-scale DQA in the transverse-field Ising model, such as particular instances of perturbed Hamming weight oracle problems, where a diabatic cascade can be formed\cite{Muthukrishnan2016}. In these cases however a classical algorithm can effectively recover the speedup since no quantum tunneling occurs in those examples. Only a single example is known where stoquastic DQA is expected to outperform QMC (and any classical algorithm): the oracular glued-tree problem studied by Childs \textit{et al} and Somma \textit{et al} reveals a limitation on the ability of classical algorithms to simulate quantum processes\cite{Childs2003, Somma2012}.

DQA imposes a new set of challenges with regard to the necessity of anti-crossings in the energy spectrum, which, similarly to the case of AQA, requires insight into their formation\cite{Choi2020, Choi2021}. Suitable anti-crossings are required in DQA, and thus for the protocol to be universally applicable, a method of guaranteeing they exist for arbitrary problems is required, among other requirements such as a large spectral separation between the first and second excited states. Furthermore, considering that QMC algorithms generally fail due to so-called sign problems, which are intricately related to the notion of non-stoquasticity of the Hamiltonian being sampled\cite{Troyer2005, Marvian2019, Gupta2019}, it is believed that using DQA on problems of a non-stoquastic nature could lead to demonstrable quantum-enabled speedups\cite{Crosson2020}. Furthermore, it has recently been argued that non-stoquasticity is an essential requirement of an annealing Hamiltonian for demonstrating such speedups\cite{Choi2021}. As of yet, there is little work on the experimental implementation of hardware that provides non-stoquastic terms\cite{Ozfidan2020}, and further, the hardware required to implement tuneable non-stoquastic two-local interaction terms\cite{Kerman2019} is currently under development and is not expected to be available for some time. As such, little is known about how a quantum annealing processor will perform with a diabatic protocol, and it is expected that the environment will play a greater role in determining the feasibility of the method\cite{Crosson2020}.

As a means to address this using near-term novel quantum annealing hardware\cite{Weber2017,Rosenberg2017}, we propose a heuristic method for programming a stoquastic quantum annealer in the diabatic mode of operation, which we call the locally suppressed transverse-field (LSTF) protocol for DQA. We show that in addition to its relevance to classical and quantum optimization, this method can be used to study the role of the environment in DQA. To this end, we first introduce a simple model of magnetic frustration to show that quantum tunneling effects can be leveraged to create a small avoided level crossing. We then introduce a means of creating an additional avoided crossing via customized annealing schedules\cite{Consani_2020, Khezri2021a, Khezri2021b} that allows us to sketch a heuristic algorithm for finding lower energy eigenstates of the problem than in AQA under certain circumstances. We also provide a statistical analysis of randomly generated examples to demonstrate the performance of LSTF-DQA and we analyze in detail a single instance to discuss the consequences of its application. In the final section, we present closed- and open-system dynamics simulations using the adiabatic master equation\cite{Albash2015a} (AME) with our frustrated magnet model where the AQA and DQA protocols are compared. We show that relaxation effects play a more important role in LSTF-DQA than in AQA. Through a detailed analysis of the eigenstates of our model and the Lindblad operators, we show that, under specific circumstances, LSTF-DQA can be used to ensure that the rate of relaxation in a larger system is determined by the target qubit. This property provides new opportunities to characterize the impact of the environment in a large processor based on annealing measurements\cite{Munoz-Bauza2019}, as an alternative to conventional dynamical methods\cite{Blais2004, Wallraff2007, Bylander2011}.

\section{Magnetic Frustration and Suppression of Tunneling}

Before we introduce LSTF-DQA formally, we will first highlight the physics underlying the methodology, to show that qubits that are frustrated due to an unfavorable local field participate in tunneling phenomena. To this end, we first look at a simple model of magnetic frustration, involving two qubits in the transverse-field model of quantum annealing,
\begin{equation}
    \hat{H}(s) = (1-s)\sum_{i \in \{1,2\}}h^x_i\hat{\sigma}_i^x + s\left[\sum_{i \in \{1,2\}}h_i^z\hat{\sigma}_i^z + J^{zz}\hat{\sigma}_1^z\hat{\sigma}_2^z\right]
    \label{eqn:H-2-qubit-cluster}
\end{equation}
where $s=t/t_{an}$ for $s \in [0, 1]$ is the dimensionless annealing parameter and we are explicitly using linear interpolation schedules for simplicity. The Pauli operators $\hat{\sigma}^\alpha$ for $\alpha=\{x, z\}$ are defined with respect to the computational basis states $|\downarrow\,\rangle=|0\rangle$ and $|\uparrow\,\rangle=|1\rangle$, i.e. $\hat{\sigma}^z  = |\downarrow\,\rangle\langle\downarrow|-|\uparrow\,\rangle\langle\uparrow|$ and $\hat{\sigma}^x  = |\downarrow\,\rangle\langle\uparrow|+|\uparrow\,\rangle\langle\downarrow|$. This Hamiltonian is necessarily stoquastic, as there is always a basis in which all off-diagonal elements are non-positive. Here we can simply choose a $\pi$ rotation in the XY plane such that the values of $h_i^x$ are all negative to satisfy this condition. Frustration is introduced in a trivial way by choosing $J^{zz}$ to be either positive or negative, and then choosing the values of $h_i^z$ such that one of the spins is biased in an unfavorable way. Using an energy scale set by $R$, and biasing qubit 2 as $0 < h_2^z < R$, we have:
\begin{equation}
    h_1^z = 
    \begin{cases}
        R & \quad \mathrm{if } J^{zz} > 0 \\
        -R & \quad \mathrm{if } J^{zz} < 0
    \end{cases}
    \label{eqn:z1-local}
\end{equation}
for qubit 1, which creates an unfavorable condition for qubit 2 to align with its local external field, i.e. in the ground state of this problem qubit 2 will be polarized against its local field. We will now focus only on the case $J^{zz}=R$. Expressing the local field of qubit 2 as $h_2^z = Rf$ where the dimensionless quantity $f$ parametrises the frustration between low ($f \rightarrow 0$) and high ($f \rightarrow 1$), the ground and first excited states are respectively,
\begin{equation}
\begin{split}
    E_{\downarrow \uparrow} &= R(f-2)\\
    E_{\downarrow \downarrow} &= -Rf
\end{split}
\end{equation}
where the gap at $s=1$ will be $\Delta E_1(s=1) \equiv E_{\downarrow \downarrow} - E_{\downarrow \uparrow} = 2R(1-f)$. When the energy scales of the transverse and longitudinal parts are matched (i.e. $h^x_{1,2} = R$), there will always be values of $s$ where the energy gap between the ground and first excited states is less than $\Delta E(s=1)$ for any $0 < f < 1$. We denote the location of the minimum energy gap in the energy spectrum as $s_*$. For a frustrated system, the energy gap at $s=s_*$ will become very small compared to $R$ as $f \rightarrow 1$. Such behavior is indicative of suppressed quantum tunneling as $f$ is increased\cite{Amin2009, King2016}.
\begin{figure}[t]
\includegraphics[scale=0.45]{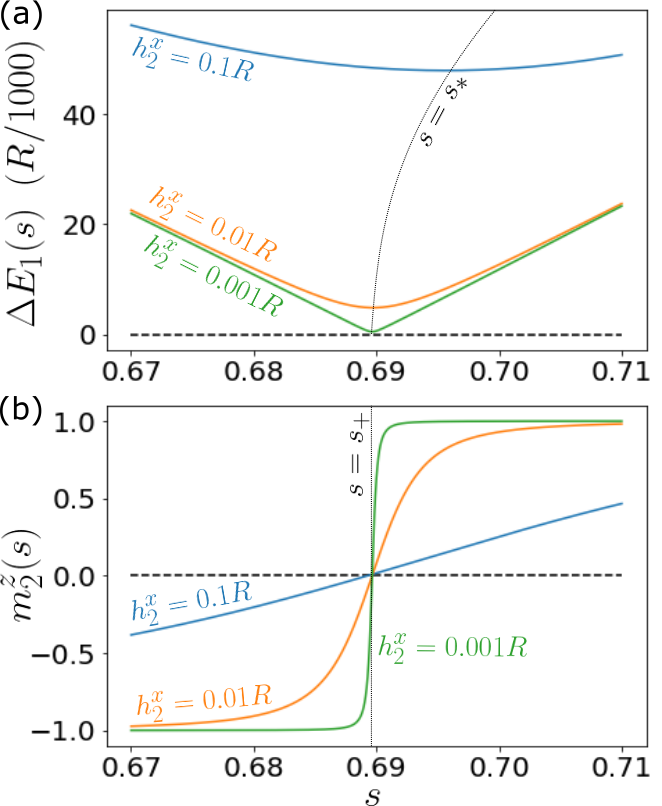}
\caption{\label{fig:spectrum} The energy gap between the ground and first excited state of Eq. \ref{eqn:H-2-qubit-cluster} as a function of $s$ is shown in (a), focused around the point $s_*$ for selected values of $h^x_2 = 0.001R$, $0.01R$ and $0.1R$, and where $f=0.8$, $J^{zz} = R$ and $h^z_1=R$ as in Eq. (\ref{eqn:z1-local}). The Z magnetization expectation value $m^z_2$ is shown in (b) for the same parameter values, and the point $s_+$ is denoted as the vertical dotted line. Note that $s_*$ converges towards $s_+$ as $h_2^x$ is decreased.}
\end{figure}

We may further control the suppression of quantum tunneling by decreasing the transverse field of qubit 2, which we call the target qubit. Figure \ref{fig:spectrum}(a) shows the energy gap between the ground and first excited state in the vicinity of $s_*$ where $\Delta E_1(s) \equiv E_1(s) - E_0(s)$ in units of $R$, solved by numerically diagonalizing Eq. (\ref{eqn:H-2-qubit-cluster}) for a selection of values $h^x_2 \ll R$ and where $h^x_1 = R$ and $f=0.8$. From this we see that it is possible to reduce the energy gap associated with the avoided-crossing by reducing the value of the transverse field applied to a single qubit, independently of the other qubit, significantly increasing the computational complexity for the adiabatic algorithm. Now consider the expectation value of the Z magnetization of the target qubit in the instantaneous ground state, $m^z_{2}(s) = \langle E_0(s)| \hat{\sigma}^z_{2} | E_0(s) \rangle$. Figure \ref{fig:spectrum}(b) shows the target qubit magnetization $m^z_{2}(s)$ solved in the same range of $s$ values, where the magnetization value crosses zero with increasing sharpness as $h^x_2$ is decreased. The point at which $m^z_{2}(s)=0$ is denoted as $s_+$ and is shown in the figure. At the point $s_+$, the target qubit is in a superposition state, indicating that the ground state of the system must be delocalized. This suggests that there are two local minima that are competing to become the global ground state. It follows from this that decreasing the $h_2^x$ term increases localization or in other words, reduces the interaction between specific minima in the potential energy, in turn increasing computational complexity for the adiabatic algorithm. It is worth pointing out that in general $s_+ \neq s_*$, and that $s_+$ and $s_*$ converge as $h^x_2 \rightarrow 0$, an observation that we discuss briefly in Appendix \ref{app:sstarsplus}.

To shed some light on the mechanism underlying the zero-crossing of the Z magnetization, we consider the semiclassical potential in the spin-coherent path-integral formalism for spin-1/2 particles\cite{Klauder1979}. We express an arbitrary two-qubit state as a tensor product of coherent states $|\theta_1,\theta_2\rangle = |\theta_1 \rangle \otimes |\theta_2 \rangle$,
\begin{equation}
    |\theta_{1,2}\rangle = \cos\left(\frac{\theta_{1,2}}{2}\right)|0\rangle + \sin\left(\frac{\theta_{1,2}}{2}\right)|1\rangle
\end{equation}
where $\theta_{1,2}$ represent the magnetization angles with respect to the down state in the XZ plane for qubits 1 and 2 respectively. We have neglected the Y magnetization component since we analyze the evolution of the potential in a closed-system setting, in the quasi-static limit where the adiabatic theorem is fully satisfied. The magnetization expectation vectors of the qubits do not have a Y component in this case as there is no external Y field and no dynamical effect that would induce a rotation out of the XZ plane. We then look at the expectation value of the energy as a function of both $\theta_1$ and $\theta_2$ with respect to the spin coherent state, which represents the semiclassical potential energy
\begin{equation}
    V(s,\theta_1,\theta_2) = \langle \theta_1,\theta_2 | \hat{H}(s) | \theta_1,\theta_2 \rangle .
    \label{eqn:V-2-qubit-cluster}
\end{equation}
Using Eq. (\ref{eqn:V-2-qubit-cluster}), we can find global and local potential minima which are associated with specific magnetization angles of both qubits. As a measure of fidelity of the semiclassical approximation, we look at the tracenorm distance as a function of $\theta_1$ and $\theta_2$\cite{Muthukrishnan2016}
\begin{equation}
    D(s,\theta_1,\theta_2) = \sqrt{1 - |\langle\theta_1, \theta_2| E_0(s) \rangle|^2}
    \label{eqn:D-2-qubit-cluster}
\end{equation}
where $|E_0(s)\rangle$ is the instantaneous ground state of Eq. (\ref{eqn:H-2-qubit-cluster}) obtained from numerical diagonalization. 
\begin{figure}[t]
\includegraphics[scale=0.45]{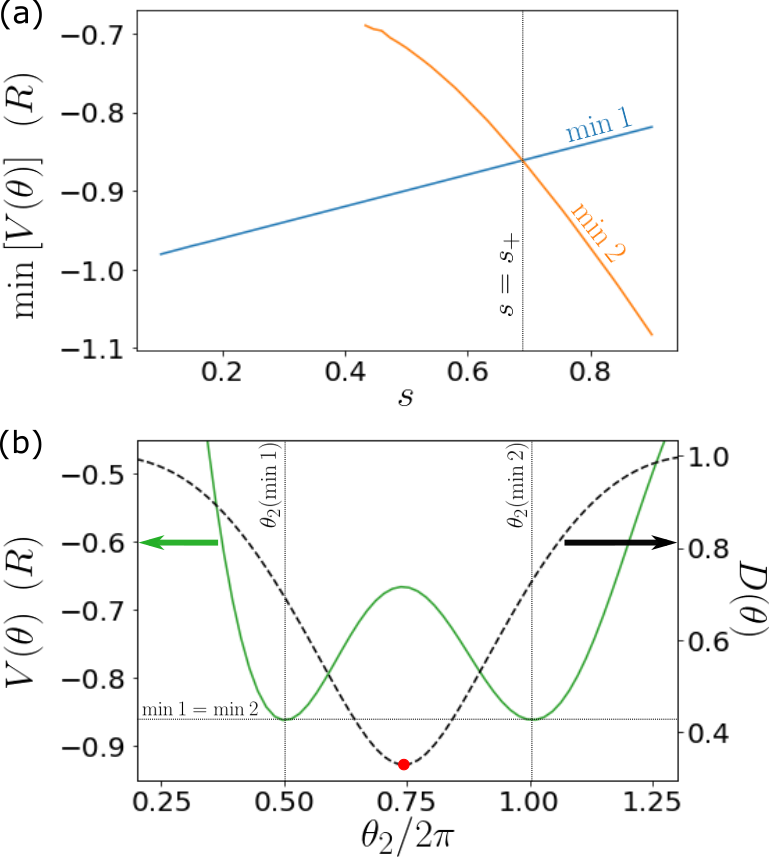}
\caption{\label{fig:potential} The energy of the two local minima as a function of $s$ are shown in (a), where the vertical line, corresponding to the point $s_+$, shows where the local minima become degenerate. (b) shows the potential energy along the line $\theta$ (see Appendix \ref{app:semiclassical} and Fig. \ref{fig:potential-3d}(a)) as a function of $\theta_2$, at the point $s=s_+$ for $h^x_2=0.01R$. The black dotted line shows the value of $D$ along the same path, which shows that the best description of the state lies in between the two minima. The global minimum of the tracenorm distance $D$ (Eq. (\ref{eqn:D-2-qubit-cluster})) is indicated by the red dot, and the values of the angle $\theta_2$ are indicated at min 1 and min 2.}
\end{figure}

We solve Eqs. (\ref{eqn:V-2-qubit-cluster}) and (\ref{eqn:D-2-qubit-cluster}) on a fine grid covering all angle combinations $\theta_{1,2}$ using $f=0.8$ and $h^x_2 = 0.01R$. We then find the local minima of the semiclassical potential numerically. Details of this analysis are provided in Appendix \ref{app:semiclassical}. Figure \ref{fig:potential}(a) shows the energy of the two local minima as a function of $s$. Early in the anneal, for $s < 0.41$, there is a single minimum in the potential, denoted min 1, which is expected as the transverse part of the Hamiltonian dominates. A local minimum in the potential, denoted min 2, then appears at a high energy and subsequently approaches the global minimum as $s$ is increased. At the point $s_+$, the two local minima become degenerate, where $m^z_{2}(s)$ is zero. For $s>s_+$, min 2 becomes lower in energy than min 1, and it therefore becomes the new global minimum. Figure \ref{fig:potential}(b) shows the potential energy along the line that connects the two minima, which we denote $\theta (s, \theta_1, \theta_2)$, as a function of $\theta_2$ at the point $s=s_+$, where the locations of min 1 and 2 are highlighted. The values of $\theta_2$ at min 1 and 2 correspond to the target qubit being in the down and up spin states, i.e. $\theta_2= \pi$ and $\theta_2=2\pi$ respectively. Now looking at the global minimum of Eq. (\ref{eqn:D-2-qubit-cluster}) along the line $\theta$ (denoted by the red dot), the value of the angle $\theta_2$ corresponds very closely to the target qubit being in an equal superposition state, i.e $\theta_2=3\pi/2$, which yields a Z expectation value $m^z_{2}(s_+) = 0$. Furthermore, the value of the tracenorm distance $D$ at $s=s_+$ is far from zero, indicating that a semiclassical description of the system ground state at this point is inaccurate. In addition, the observation that the best classical description of the system ground state at $s=s_+$ lies between the two local minima of the semiclassical potential is an indication that the state is delocalized across both potential wells. We share plots of the full potential landscape and computed values of $\theta$ in Appendix \ref{app:semiclassical}.

We can see from both Figs \ref{fig:potential}(a) and \ref{fig:potential}(b) that the change in the global potential minimum occurs discontinuously in $s$, i.e. there is no classical path for a particle to move from min 1 to min 2 during the evolution of the algorithm. In the absence of a thermal bath, this indicates that a tunneling event must occur near $s_+$, characterized by a flip in the Z component of the magnetization expectation value. The value of $h^x_2$ controls the width and height of the potential barrier, and therefore the interaction between local minima in the potential. This is also indicated by the width of the $s$ interval in which the Z component of the qubit magnetization expectation value flips. This width becomes smaller as $h^x_2$ decreases, as seen in Fig. \ref{fig:spectrum}(b), indicating that the target qubit is increasingly localized in either the up ($m^z_{2}=+1$) or down ($m^z_{2}=-1$) state before and after $s_+$ respectively. In the limit $h^x_2 \rightarrow 0$, the interaction between the local minima is switched off, which leads to an energy crossing\cite{Altshuler2010}. We expect that in large-scale problems, magnetic frustration caused by unfavorable local Z fields lead to multiple competing local minima\cite{King2016}, which in turn can be individually controlled using local transverse fields. An explicit example of this is discussed in the context of the Strong-Weak cluster problem\cite{Albash2021}, where it is shown that a diagonal catalyst can remove competing local minima and thereby soften the exponentially closing energy gap of the problem. Finally we note that commercial annealers do not yet offer the ability to control the transverse fields of the qubits independently\cite{Boothby2020}. The need for increased flexibility of the annealing schedules, including independent control of the transverse fields, has been widely recognized (see \cite{Khezri2021a} and references therein), putting such capabilities on the roadmap for future devices\cite{Rosenberg2017, Novikov2019, Kerman2019, Grover2020, Khezri2021b}.

\section{Locally Suppressed Transverse-Field DQA}

We now describe how the controlled suppression of quantum tunneling in a frustrated qubit can be exploited for creating DQA protocols. In this section we will focus on the computational application of this effect, where it is used heuristically. We have seen so far in a simple example that the expectation of the Z magnetization of a frustrated qubit will cross zero as $s$ increases from 0 to 1. This situation corresponds to the existence of a delocalized state at $s=s_+$ due to quantum tunneling, which can be effectively suppressed by reducing the X field associated with that qubit, in turn creating a very small avoided crossing. However a way of reliably creating a second avoided crossing is required to realize the simplest case DQA protocol, that is where a shortcut to the final ground state via the first excited state exists due to the existence of two avoided crossings at different values of $s$.

Consider a quantum annealer that provides individual control over the X local fields of each qubit. For an arbitrary problem graph $G$, we express the driver Hamiltonian as
\begin{equation}
    \hat{H}_D(s) = \sum_{i\in V(G)} a_i(s) h^x_i\hat{\sigma}_i^x
    \label{eqn:H-driver-custom}
\end{equation}
where the functions $a_i(s)$ are the driver schedules, now unique to each qubit. $V(G)$ is the set of vertices of $G$. We express the problem Hamiltonian as
\begin{equation}
    \hat{H}_P(s) = \sum_{i\in V(G)} b_i(s) h_i^z\hat{\sigma}_i^z + \sum_{ij \in E(G)} b_{ij}(s) J^{zz}_{ij} \hat{\sigma}_i^z\hat{\sigma}_j^z
    \label{eqn:H-problem-custom}
\end{equation}
where the functions $b_i(s)$ and $b_{ij}(s)$ are the problem schedules, and $E(G)$ is the set of edges of $G$ associated with each 2-local coupling term. In the following we will assume that $h^x_i = R\,\,\,\, \forall i \in V(G)$ unless stated otherwise.

\begin{figure}[t]
\includegraphics[scale=0.45]{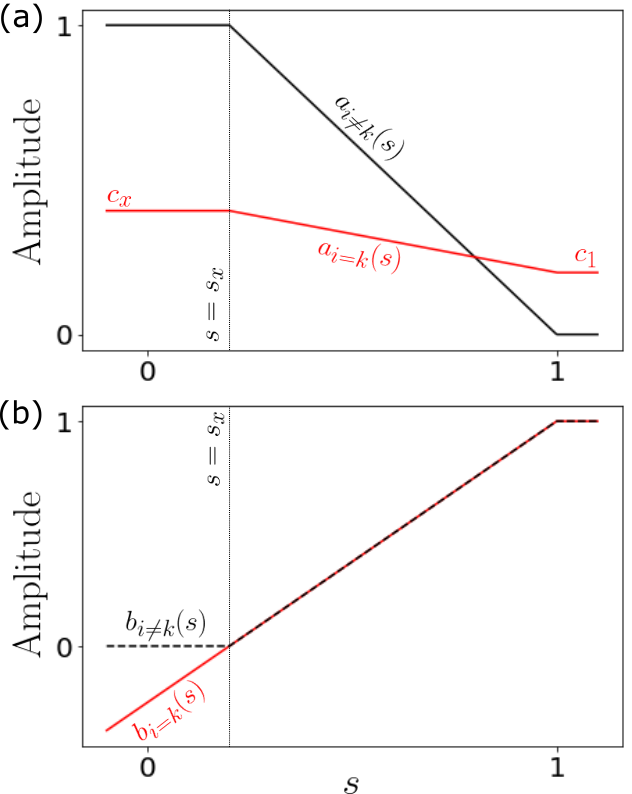}
\caption{\label{fig:dqa-schedules} The DQA schedules as a function of $s$ for the transverse part of the Hamiltonian are shown in (a), corresponding to Eqs. (\ref{eqn:X-sched-i}) and (\ref{eqn:X-sched-k}). Arbitrary values of $c_x$ and $c_1$ are shown for illustration. The DQA schedules for the longitudinal part of the Hamiltonian are shown in (b), corresponding to Eqs. (\ref{eqn:Z-sched-i}) and (\ref{eqn:Z-sched-k}).}
\end{figure}

We can trivially make an additional avoided crossing by creating a quasi-degeneracy in the energy spectrum of the driver Hamiltonian Eq. (\ref{eqn:H-driver-custom}). We define our schedules as follows, including a new parameter $s_x$ such that $0 \leq s_x \leq 1$ which defines the value of $s$ at which the additional avoided crossing occurs. We select one qubit $k$ from $V(G)$, which we refer to as the target qubit. The driver schedules as then defined as
\begin{equation}
    a_{i \neq k}(s) =
    \begin{cases}
    1       & \quad \text{if } s < s_x \\
    1 - \frac{s - s_x}{1 - s_x}  & \quad \text{if } s_x \leq s \leq 1
    \end{cases}
    \label{eqn:X-sched-i}
\end{equation}
and
\begin{equation}
    a_{i = k}(s) =
    \begin{cases}
    c_x       & \quad \text{if } s < s_x \\
    c_x + (c_1 - c_x)\frac{s - s_x}{1 - s_x}  & \quad \text{if } s_x \leq s \leq 1
    \end{cases}
    \label{eqn:X-sched-k}
\end{equation}
where $c_x$ and $c_1$ are parameters that control the amplitude of the X field of the target qubit $k$. Figure \ref{fig:dqa-schedules}(a) shows these schedules as a function of $s$ and the effect of their parameters. We see that in the case $i \neq k$ (Eq. (\ref{eqn:X-sched-i})) the schedules are the same as the usual linear function, excepting that the interpolation starts at $s_x$. The schedule of Eq. (\ref{eqn:X-sched-k}) is a new feature, where the parameter $c_x$ scales down the value of $h^x_k$. $c_1$ determines the value of $h^x_k$ at $s=1$. The parameters $c_x$ and $c_1$ control the size of the energy minima at $s=s_x$ and $s=s_*$ respectively. These are envisaged to be particularly useful for implementing multi-qubit interference experiments\cite{Munoz-Bauza2019, Karanikolas2020}, which could serve as a sensitive probe of coherence, a point we will revisit in future work. For simplicity, we assume $c_x = c_1 = 0$ for the remainder of this work. This choice closes the energy gaps at $s=s_x$ and $s=s_*$, resulting in level-crossings at these points. However it must be emphasized that in practice it might not be possible to completely eliminate the transverse field of a given qubit, depending on the hardware, but it can be made negligible in the flux qubit\cite{Orlando1999}.

The problem schedules are defined as
\begin{equation}
    b_{i \neq k}(s) =
    \begin{cases}
    0       & \quad \text{if } s < s_x \\
    \frac{s - s_x}{1 - s_x}  & \quad \text{if } s_x \leq s \leq 1
    \end{cases}
    \label{eqn:Z-sched-i}
\end{equation}
and
\begin{equation}
    b_{i = k}(s) =
    \begin{cases}
    \frac{s - s_x}{1 - s_x}       & \quad \text{if } 0 \leq s \leq 1
    \end{cases}
    \label{eqn:Z-sched-k}
\end{equation}
where again we see that for $i \neq k$ the schedules are the same linear function with a delayed start to the interpolation, as shown in Fig. \ref{fig:dqa-schedules}(b). The schedule of Eq. (\ref{eqn:Z-sched-k}) takes the same form as those in Eq. (\ref{eqn:Z-sched-i}), where the only difference is that the former is not restricted to a constant in the range $s < s_x$. This results in a zero crossing of this function at $s=s_x$, hence changing the sign of $h^z_k$ at that point. The result of these schedules is to create an avoided crossing at $s=s_x$ with a gap of size $2c_xR$ for $c_x \ll 1$. It is trivial to see that this creates a true avoided crossing, as the value of $b_k(s)h^z_k$ changes sign about $s_x$, inducing a zero crossing in the expectation of $\hat{\sigma}^z_k$, or in other words, the creation of a delocalized state at $s=s_x$. Now for our target qubit $k$, if the point $s=s_+$ exists, i.e. if qubit $k$ participates in the delocalized character of the ground state at that point, then we will now have two very small gaps between the ground and first excited states, at $s=s_x$ and $s=s_+$. Note that the schedules $b_{ij}(s)$ in the last term of Eq. (\ref{eqn:H-problem-custom}) take the same form as Eq. (\ref{eqn:Z-sched-i}). Also note that by design, $s_+ > s_x$ is always true, since the interpolation of the driver and problem Hamiltonians always starts at $s=s_x$.

We now propose the following heuristic algorithm for finding a low-energy eigenstate, given an arbitrary large-scale problem. We assume no prior knowledge of the structure of the problem, and that the system is coherent in the DQA regime.
\begin{enumerate}
    \item Apply conventional AQA to the problem.
    \item Determine the state with lowest energy obtained.
    \item For each qubit $k$ such that $h^z_k \neq 0$:
    \begin{enumerate}
        \item Apply schedules in Eqs. (\ref{eqn:X-sched-i})-(\ref{eqn:Z-sched-k}).
        \item Update the state with lowest energy if it exists.
    \end{enumerate}
\end{enumerate}
In step 1, the AQA algorithm is used as a reference. If the considered problem has a very small gap compared to the annealing times used, one or more occurrences of excited states will be obtained in step 2. In step 3 the locally suppressed transverse-field (LSTF) DQA method is applied to each qubit in turn, using a value of $c_x=0$ such that the probability of a transition at $s=s_x$ is 1. A number of possible outcomes are envisaged in step 3. In the case that the problem does not present a small gap, then the LSTF-DQA method should never yield an eigenstate with lower energy. In the case that the problem has a \textit{single} small gap, i.e. the separation with higher energy eigenstates is large, then the LSTF-DQA method should consistently yield an eigenstate with lower energy. If the separation with higher energy eigenstates is small, and an energy minimum between such states occurs some time \textit{after} the lowest energy minimum gap between the ground and first excited state, i.e. argmin$\left[E_2(s) - E_1(s)\right] > s_*$, then the DQA protocol has better chances of yielding lower energy eigenstates. On the other hand, if argmin$\left[E_2(s) - E_1(s)\right] < s_*$, then it is possible that the DQA protocol performs as badly or worse than the AQA protocol, due to transitions to higher eigenstates. Note that although this algorithm is entirely heuristic and provides no guarantee of better results than in the adiabatic case, the $O(N)$ scaling of step 3 makes it attractive to use as an alternative to see if a lower energy eigenstate can be obtained.

\subsection{Statistical Analysis}

To estimate the performance of the LSTF-DQA technique and the algorithm proposed above, we solve a series of randomly generated 7-qubit problem instances using the Schr\"{o}dinger equation (SE) solver of the \texttt{python} package \texttt{qutip}\cite{Johansson2012, Johansson2013} with a set annealing time $t_{an} = 100$ ns and $R = 1$ GHz. To limit the total number of random graphs, we generate a set of ten non-isomorphic connected graphs each with instances comprising 6, 8, 10, 12, 14 and 16 edges, providing a total of 60 unique graphs of increasing connectivity (see Appendix \ref{app:random-instance-graphs} for details). To generate the local Z-fields $h^z_i$ we draw values from identical and independent Gaussian distributions with a variance of 1 and mean of 0, and then normalize them such that the largest field always has a magnitude of $R$. We then set all the values of $J^{zz}_{ij}$ associated with the edges of the graphs to $-0.5 R$ to promote the occurrence of frustrated qubits. For each set of graphs with a given connectivity, we consider 500 samples of the set of seven local Z-field values initialized on a randomly selected graph from the set of ten. For each sample, the generated AQA problem is considered `small gap' (SG) if the condition $\Delta E_1(s_*) \leq 1/2\pi$ GHz is satisfied. The AQA problem is considered `large gap' (LG) otherwise. We could further say that SG problems are considered 'hard' and LG problems 'easy', but this is not the case for DQA and thus we class problems more broadly as SG and LG here. Given the evolution time $t_{an} = 100$ ns, the heuristic adiabatic condition set by the chosen minimum gap threshold for hardness can be expressed approximately as $t_{an} \geq 4000/\Delta E_1^2(s_*)$, which is consistent with the commonly used expression $t_{an} \gg \Delta E_1^{-2}(s_*)$. We then solve the SE applied to both the AQA algorithm and the LSTF-DQA technique applied to each qubit in turn. In all sample runs, the LSTF-DQA schedule parameters were $s_x=0.2$ and $c_x=c_1=0$. For each sample, we calculate the energy residual
\begin{equation}
    E_\mathrm{res} = \langle E_f| \hat{H}_P(1) |E_f \rangle - E_0
    \label{eqn:Eres}
\end{equation}
where $E_0$ is the ground state energy of $\hat{H}_P(1)$ and $|E_f \rangle$ is the final state obtained from the SE at $s=1$, for both the AQA protocol, and for each qubit under the LSTF-DQA protocol. Similarly, we also calculate the probability of obtaining the ground state as $|\langle E_f| E_0 \rangle|^2$ for each case.
\begin{table}[b]
\caption{\label{tab:ran-stat} Summary of the performance of the LSTF-DQA protocol for the different groups of graph connectivities used. Each column represents statistics covering 1000 samples and 20 graphs. A DQA win is determined when the smallest energy residual obtained from Eq. (\ref{eqn:Eres}) is less than that obtained for the equivalent AQA problem.}
\begin{ruledtabular}
\begin{tabular}{cccc}
\textrm{Instance Edges}&
\textrm{6 and 8}&
\textrm{10 and 12}&
\textrm{14 and 16}\\
\colrule
\textrm{\parbox[t]{3cm}{DQA Wins}} & 61.7 \% & 60.6 \% & 54.5 \%\\
\textrm{\parbox[t]{3cm}{Small Gap Problems}} & 27.4 \% & 16.9 \% & 8.5 \%\\
\textrm{\parbox[t]{3cm}{Small Gap DQA Wins}} & 73.3 \% & 77.7 \% & 97.7 \%\\
\end{tabular}
\end{ruledtabular}
\end{table}
\begin{figure*}
\includegraphics[scale=0.45]{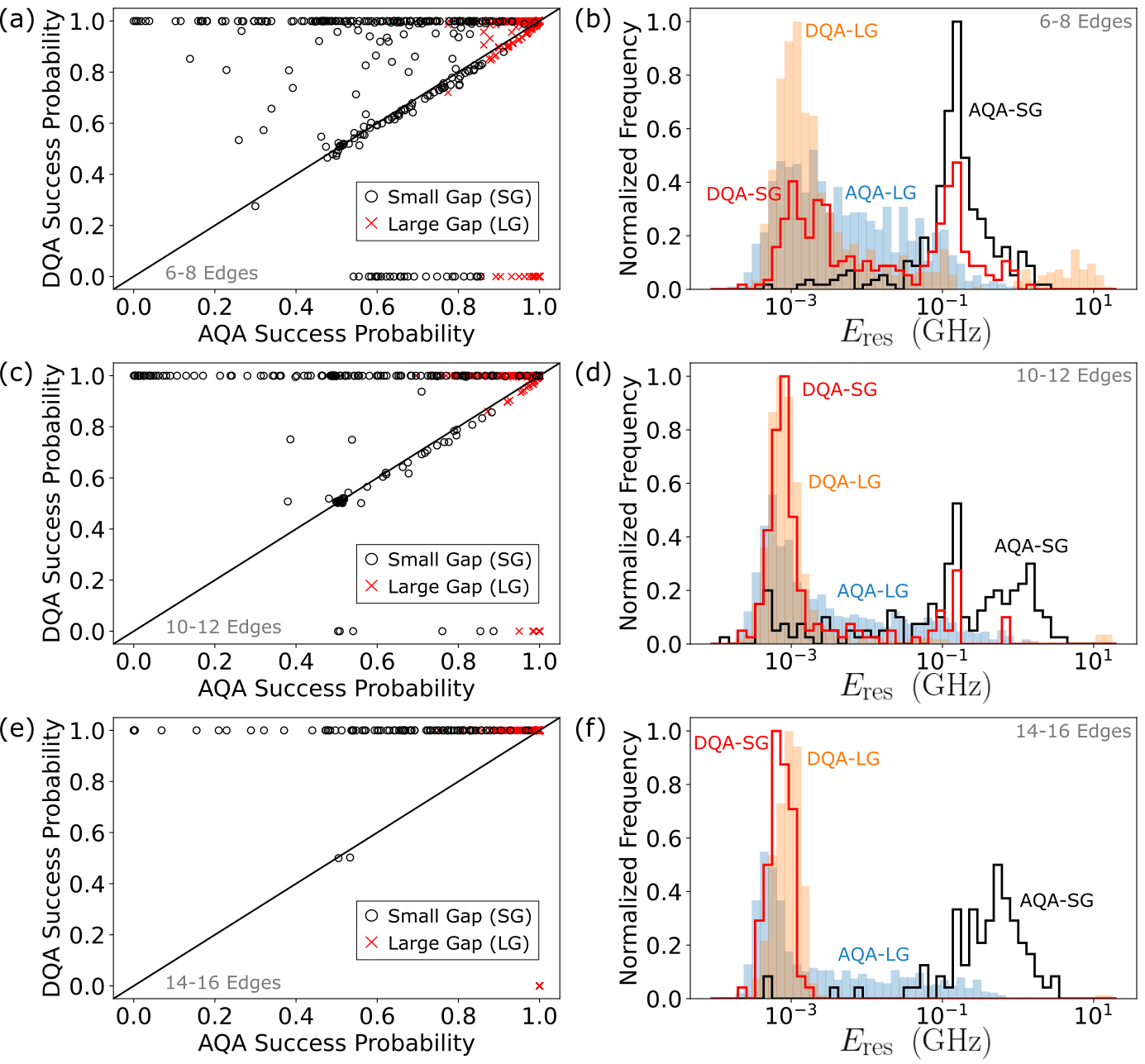}
\caption{\label{fig:ran-stat-all} Summary of success probabilities (a), (c), (e), and energy residual histograms (b), (d), (f), of all the randomly generated 7-qubit problems. The rows (a)-(b), (c)-(d) and (e)-(f) correspond to the instance groups with 6-8, 10-12 and 14-16 edges respectively (annotated in gray font). In the plots (a), (c) and (e), the black line denotes equal success probabilities, the black circles indicate SG instance samples and the red crosses LG instance samples. In the histograms (b), (d) and (f), the energy residuals are represented as four distinct groups according to whether the AQA or LSTF-DQA was used, and the classification of the problem, small gap (SG) or large gap (LG), determined by the minimum gap threshold. A problem is considered to have a SG when $\Delta E_1(s_*) \leq 1/2\pi$ GHz in the AQA protocol, and LG otherwise. The bin frequencies corresponding to SG and LG samples in the histograms were normalized for the sake of clarity. Refer to Table \ref{tab:ran-stat} for the true proportion of SG and LG problems.}
\end{figure*}

Table \ref{tab:ran-stat} summarizes the overall performance of LSTF-DQA algorithm for each set of graph connectivities and where we have lumped the graph sets into groups of two for legibility. A win for LSTF-DQA is determined when the smallest energy residual obtained from Eq. (\ref{eqn:Eres}) is less than that obtained for the equivalent AQA problem. From these results we see that in general, LSTF-DQA performs better by a small margin that slightly decreases for increased graph connectivity. However, LSTF-DQA performs much better on the proportion of SG problems detected, reaching near 100\% success for the high-connectivity SG problems.

Figures \ref{fig:ran-stat-all}.(a)-(f) depict all the data obtained for both the success probability and energy residual figures of merit. The success probabilities are shown in Figs \ref{fig:ran-stat-all}.(a), \ref{fig:ran-stat-all}.(c) and \ref{fig:ran-stat-all}.(e) corresponding to the instance groups with 6-8, 10-12 and 14-16 edges respectively. The energy residuals are shown in Figs \ref{fig:ran-stat-all}.(b), \ref{fig:ran-stat-all}.(d) and \ref{fig:ran-stat-all}.(f) corresponding to the instance groups with 6-8, 10-12 and 14-16 edges respectively. In the case of the LG problems, which represent the majority of the samples, we see that the success probability is generally high for both the AQA and LSTF-DQA algorithms, represented by the red crosses, with most points lying in the top right quadrant of each graph. However there are cases where the LSTF-DQA algorithm gets trapped in higher energy states, as evidenced by the occurrence of zero success probabilities and high energy residual values near 10 GHz. These cases are particularly evident for the problem instances with the more sparse graphs comprising 6 and 8 edges, as seen in Figs \ref{fig:ran-stat-all}.(a) and \ref{fig:ran-stat-all}.(b). In the case of the SG problems, LSTF-DQA generally performs much better than the AQA algorithm, particularly for the denser graphs comprising 10, 12, 14 and 16 edges. This is made evident by the low energy-residual values obtained for the SG DQA problems compared to the SG AQA problems. In the case of the SG problems in particular, there is a clear separation between the energy-residuals obtained for LSTF-DQA and those obtained for AQA. This shows that LSTF-DQA has great potential for solving SG problems.
\begin{figure}[b]
\includegraphics[scale=0.45]{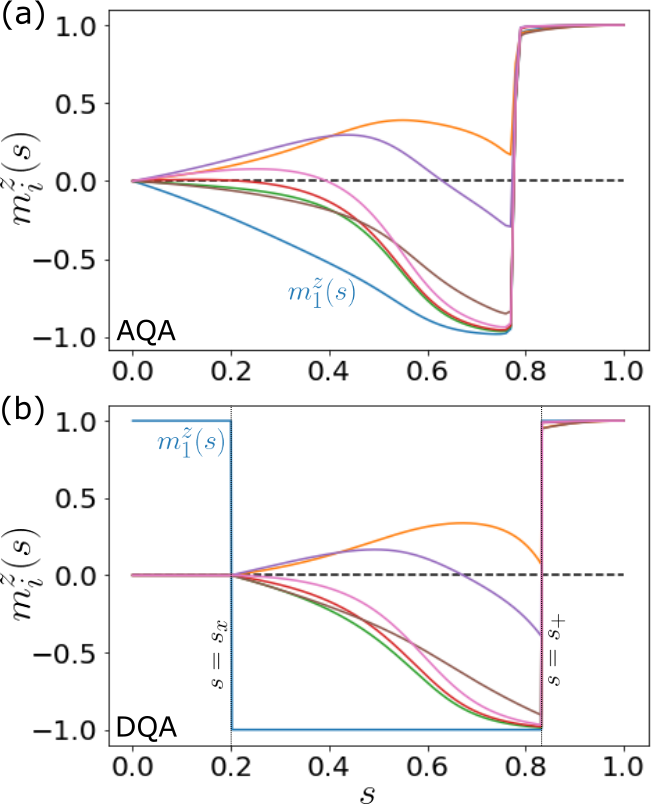}
\caption{\label{fig:random-magz} The magnetization expectation values of each qubit using (a) the AQA algorithm with linear schedules, and (b) LSTF-DQA applied to qubit 1 ($k=1$) of the graph shown in Fig. \ref{fig:random-G}. In (b) the points $s=s_x$ and $s=s_+$ are shown.}
\end{figure}

An interesting feature of the success probability plots is the gathering of many points along and just below the black line representing equal success for LSTF-DQA and AQA, particularly for the more sparse 6 and 8 edge graph instances. These correspond to problems in which the ground state is increasingly doubly degenerate as the points along this line approach 0.5 success probability. LSTF-DQA performs slightly worse for these types of problems, due to the tendency for the transverse-field suppression to cause the ground and first excited states to be highly degenerate throughout the anneal. Problem instances with degenerate ground states become less common as the graph connectivity is increased. This is because spin flips are penalized much more on average due to the increased number of edge penalties. We also notice that some points are scattered above the black line, for both LG and SG instances. In these cases, the minimum gap in the energy spectrum occurs only slightly before the end of the anneal, causing the final state to be hybridized with higher excited states at the chosen evolution rate. LSTF-DQA generally performs better in these cases as the final state obtained is closer to the true ground state than for the AQA algorithm. Finally, we notice that some hard problems have zero success rate for LSTF-DQA. These are problems similar to those that lie near the black line, in that they have nearly or completely degenerate ground states, however they have the added feature that no qubits are frustrated. In this case, LSTF-DQA does not work as intended and only ends up creating paths to higher excited states.
\begin{figure}[b]
\includegraphics[scale=0.45]{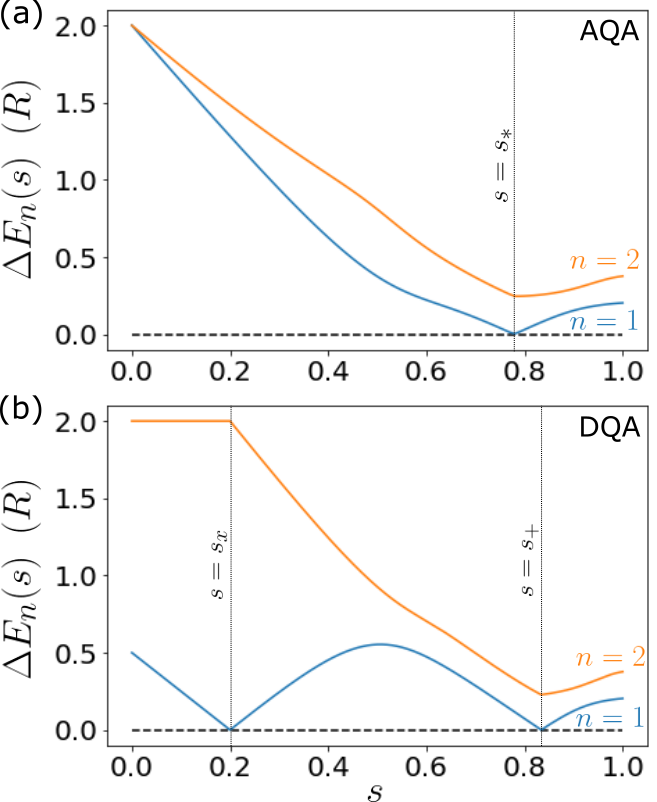}
\caption{\label{fig:random-spectrum} The spectrum of the randomly generated problem using (a), the AQA algorithm with linear schedules, and b) LSTF-DQA applied to qubit 1. In (a) the minimum gap location $s=s_*$ is shown, and in (b) the points $s=s_x$ and $s=s_+$ are shown. Only the two lowest energy gaps are shown here for clarity.}
\end{figure}

\subsection{Single Instance Analysis}

We now investigate in detail a single instance. We consider a sample problem with a graph comprising ten edges, which can be considered hard for AQA due to the presence of a phase transition. See Appendix \ref{app:random-instance-vals} for the values of $h^z_i$ and the graph used for this instance.  Figure \ref{fig:random-magz} shows the expectation values of $\hat{\sigma}^z_i$ during a normal linear annealing process (a), and using LSTF-DQA in (b), where we have selected $k=1$ for the target qubit, with $s_x=0.2$ and $c_x=0$. In Fig. \ref{fig:random-magz}.(a) we can see that most qubits have a Z expectation that crosses zero at various points in $s$, and in particular near $s=0.8$ where there is a phase transition, characterized by a sharp and drastic change in total magnetization. With this transition is associated a small gap\cite{Nishimori2016} as seen in Fig. \ref{fig:random-spectrum}.(a) at $s=s_*$, where the energy spectrum is represented as energy gaps $\Delta E_n(s) \equiv E_n(s) - E_0(s)$. The minimum gap in this instance is well below the chosen threshold for hardness, i.e. $\Delta E_1(s_*) \leq 1/2\pi$ GHz. In Fig. \ref{fig:random-magz}.(b) we show the effect of applying the schedules described in Eqs. (\ref{eqn:X-sched-i})-(\ref{eqn:Z-sched-k}) on the magnetization expectations of each qubit $m^z_i$. We can see that $m^z_1$ undergoes two sharp transitions at the points $s=s_x$ and $s=s_+$ as expected, which leads to energy crossings between the ground and first excited state. The LSTF-DQA spectrum is shown in Fig. \ref{fig:random-spectrum}(b) where the double energy crossings can clearly be identified. For the evolution time $t_{an} = 100$ ns, we obtained a success probability of less than 0.01 for the AQA algorithm, whereas for the LSTF-DQA algorithm we obtained a success probability over 0.99 for three qubits (qubits 1, 3 and 6). The best result was obtained for qubit 1, with a success probability over 0.9999. In terms of the time-to-solution (TTS) for this single run, calculated using Eq. (1) in Ref. \cite{Albash2018a} and where we use a success criterion $p_d=0.99$, these results represent a constant speedup factor on the order of 1000 times for LSTF-DQA over AQA, for which we obtained 46.7 ns and 49 $\mathrm{\mu s}$ respectively. Another observation is that some of the qubit Z magnetization expectation values undergo two zero crossings. For those qubits, two minima in the energy spectrum can be created by only reducing their associated X field. We discuss this in more detail in Appendix \ref{app:random-instance-props}.

We have demonstrated that a constant speedup is achievable using LSTF-DQA for a set of problems, as opposed to a scaling advantage, and in particular for problems considered hard for quantum annealing. Despite this it is useful to emphasize that the problems considered in this work are still easily solved classically (due to the limited number of qubits in the simulations), and further that LSTF-DQA can not be applied to qubits that have no local Z field in the problem Hamiltonian, such as quantum random-energy models\cite{Jorg2008, Jorg2009} and synthetic deceptive-cluster-loop benchmark problems\cite{Mandra2018}. However, given the performance we have demonstrated on small-scale problems, we expect that the LSTF-DQA algorithm proposed in this work will be particularly useful as an alternative annealing method for solving large-scale problems in which there is a phase transition and that feature an inhomogeneous disorder field\cite{Grass2019}, such as that discussed in the previous paragraph. The performance of LSTF-DQA at scale is an aspect we wish to revisit in future work, using scalable models that exhibit first order phase transitions. The method could also be used to elucidate the structure of a black-box problem, an application which we also wish to investigate in further work. Owing to recent work on the accurate mapping of superconducting circuit Hamiltonians to Ising Hamiltonians\cite{Consani_2020}, and efficient methods for finding physical annealing schedules that accurately reproduce the desired scheduling of the Ising terms\cite{Khezri2021a}, we anticipate that the schedules proposed in Eqs. (\ref{eqn:X-sched-i})-(\ref{eqn:Z-sched-k}), or similar variants, can be readily implemented provided the qubit design is suitable (e.g. it allows a near-zero transverse field).

\section{Dynamics Simulations}

\begin{figure}[t]
\includegraphics[scale=0.45]{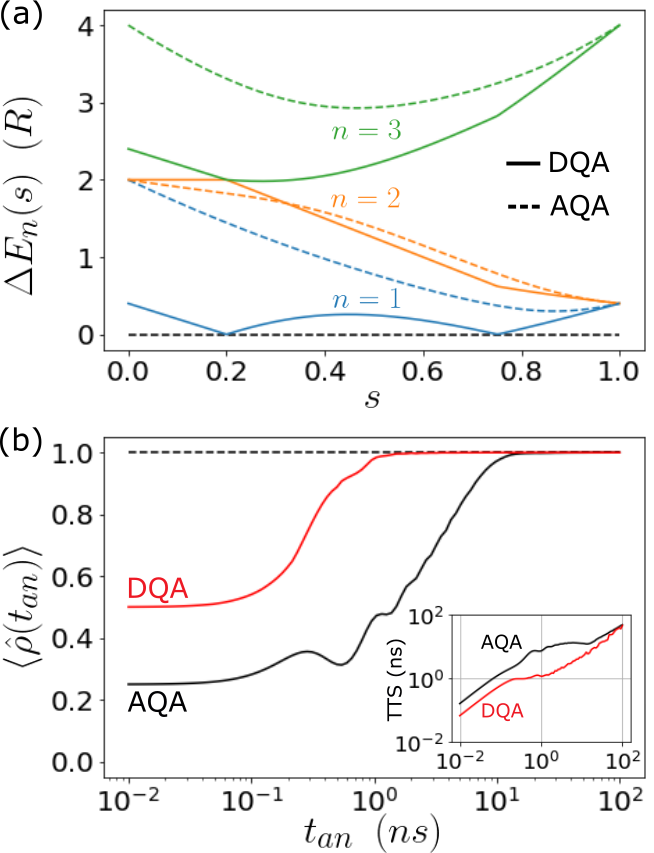}
\caption{\label{fig:aqa-dqa-cs} The complete two-qubit energy spectra in the AQA and LSTF-DQA protocols are shown in (a) as dotted and solid lines respectively. Solutions for $\langle \hat{\rho}(t_{an}) \rangle$ as a function of the total annealing time for the linear AQA and LSTF-DQA protocols are shown in (b). The inset of (b) shows the $t_{an}$ dependence of time-to-solution (TTS) for a success criterion of $p_d=0.99$. The parameters of the annealing Hamiltonian here are $R=1$ GHz, $c_x=0$ and $f=0.8$.}
\end{figure}

We anticipate that the annealing processor environment will play an equal if not more critical role in determining DQA performance compared to AQA. To assess this in the context of LSTF-DQA, we revisit our model of magnetic frustration, using the 2-qubit Hamiltonian as defined in Eq. (\ref{eqn:H-2-qubit-cluster}) and perform dynamics simulations in a closed-, and open-system setting. We use the Hamiltonian Open Quantum System Toolkit (HOQST) set of \texttt{Julia} codes for solving all master equations\cite{Chen2020}. In our simulations we set the energy scale $R=1$ GHz and frustration $f=0.8$ which yields the minimum gap $\Delta E_1(s_*) \approx 0.4$ GHz and $s_* \approx 0.85$ in the AQA case. Figure \ref{fig:aqa-dqa-cs}(a) shows all the energy gaps associated with both the AQA and LSTF-DQA instances of this Hamiltonian using these parameters. 

We first present closed-system dynamics simulations comparing both the AQA and LSTF-DQA implementations of the schedules defined in Eqs. (\ref{eqn:X-sched-i})-(\ref{eqn:Z-sched-k}). We solve numerically the von-Neumann equation for a range of annealing times $t_{an}$,
\begin{equation}
    \hbar \frac{\partial}{\partial t}\hat{\rho}(t) = -i \left[ \hat{H}(t), \hat{\rho}(t) \right]
    \label{eqn:von-neumann}
\end{equation}
where $\hat{\rho}(t)$ is the density matrix and $\hat{H}(t) = \hat{H}_D(t) + \hat{H}_P(t)$. Recall that $s = t/t_{an}$ so that $t$ runs from $t=0$ to $t=t_{an}$. The initial state is the pure ground state of the system at $s=0$, $\hat{\rho}(0) = |E_0(0)\rangle \langle E_0(0)|$. The final state of the system in this case will also be a pure state. The probability of obtaining the ground state is simply $p_0(t_{an}) = \langle E_{\downarrow \uparrow} |\hat{\rho}(t_{an})| E_{\downarrow \uparrow} \rangle$. Figure \ref{fig:aqa-dqa-cs}(b) shows the result of solving Eq. (\ref{eqn:von-neumann}) to find the probability of obtaining the problem ground state at the end of the anneal $t=t_{an}$. In the case of AQA, a high ground state probability is obtained when the heuristic adiabatic theorem\cite{Albash2018} is well satisfied $t_{an} \gg 1/\Delta_1(s_*)^2$. The equivalent problem in the LSTF-DQA protocol is solved using anneal durations an order of magnitude lower. The inset of Fig. \ref{fig:aqa-dqa-cs}(b) shows the $t_{an}$ dependence of the time-to-solution (TTS) for a single run, calculated using Eq. (1) in Ref. \cite{Albash2018a}, and where we use a success criterion $p_d=0.99$. Notice in Fig. \ref{fig:aqa-dqa-cs}(a) that the separation between the ground and second excited states for $s<s_x$ and $s>s_+$ and the separation between first and second excited states for $s_x < s < s_+$ in the DQA protocol is larger than the separation between the ground and first excited in the AQA protocol. This, combined with the fact that the transition to the first excited state at $s_x$ and back to the ground state at $s_+$ occurs at any timescale due to our choice of $c_x=0$, shows that the adiabatic theorem is satisfied at shorter timescales than with the AQA protocol.

We now turn to open-system dynamics to assess the impact of the environment. As the parameter space for open-systems is very large, we restrict ourselves to the simplest cases, saving more detailed studies for future work. The first simplification we apply is to use $c_x=0$ throughout our simulations. As we have shown, in this limit there is no interaction between local minima in the semiclassical potential, and thus energy-level crossings are created between the ground and first excited states. This allows us to focus on the use of an adiabatic master equation (AME) for open-system dynamics since there is in principle no diabatic transition between the ground and first excited state\cite{Dziarmaga2010}. As a second simplification, we focus on two limiting cases of uncorrelated system-bath couplings for a given bath. We only consider the cases where an Ohmic bath is coupled equally to either the X or Z degrees of freedom of Eq. (\ref{eqn:H-2-qubit-cluster}), which amounts to considering the effects of relaxation and dephasing respectively.
\begin{figure}[t]
\includegraphics[scale=0.45]{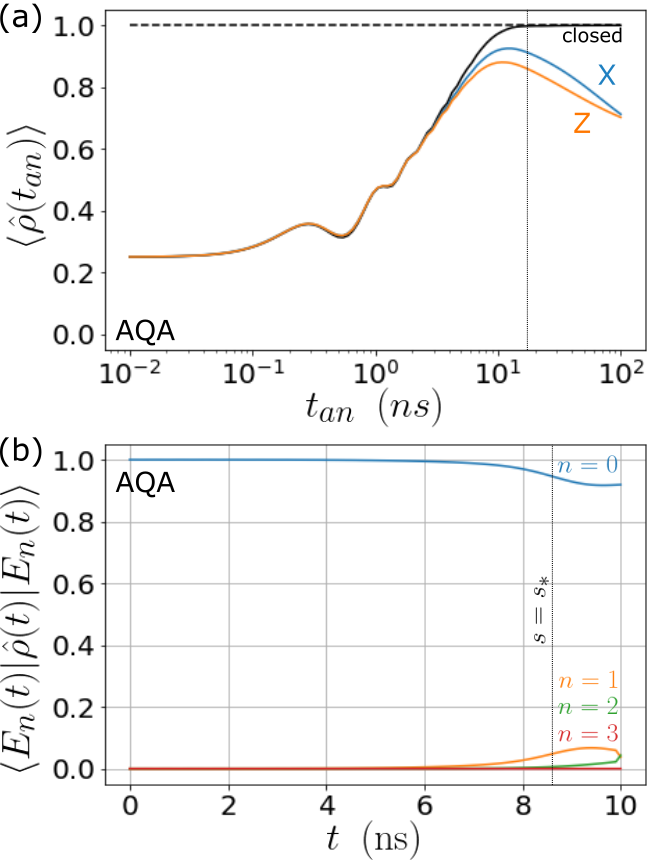}
\caption{\label{fig:aqa-xz} The probability of obtaining the ground state at the end of an anneal for the AQA protocol are shown in (a), calculated by solving the AME for the Ohmic bath coupled to the X (blue line) and Z (orange line) degrees of freedom. The closed system result is shown as the black line for reference and the vertical dotted line shows the time at which the dynamics are quasi-adiabatic. In (b) are shown the overlaps of all the instantaneous eigenstates of the system Hamiltonian with the density matrix obtained for $t_{an} = 10$ ns, in the case of the bath coupled to the X degrees of freedom. The location of the minimum gap $s_*$ is shown, as well as the indices of the states.}
\end{figure}
\begin{figure}[t]
\includegraphics[scale=0.45]{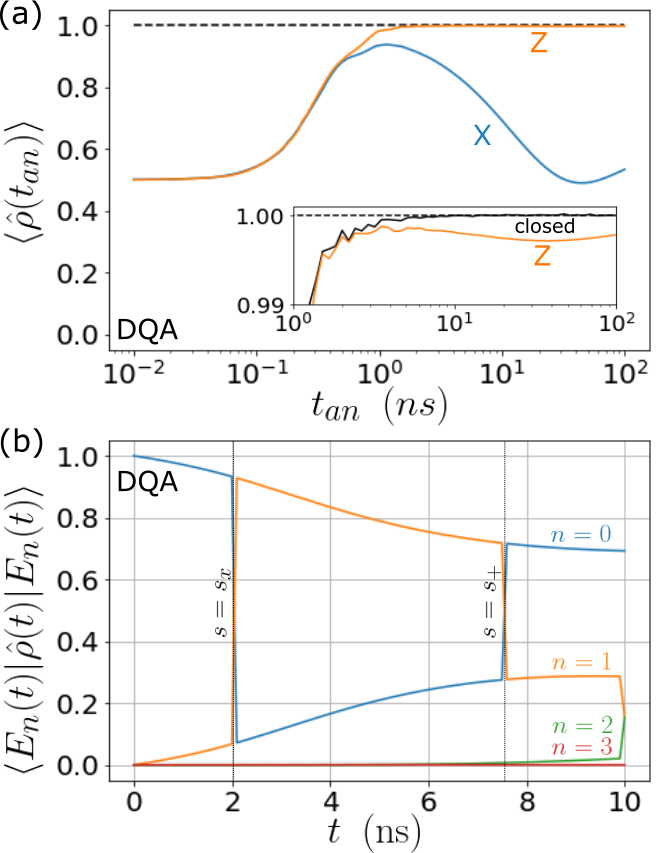}
\caption{\label{fig:dqa-xz} The probability of obtaining the ground state at the end of an anneal for the DQA protocol are shown in (a), calculated by solving the AME for the Ohmic bath coupled to the X (blue) and Z (orange) degrees of freedom. The inset shows a closeup of the closed system result (black solid line) and the Z-coupled bath result. In (b) are shown the overlaps of all the instantaneous eigenstates of the system Hamiltonian with the density matrix obtained for $t_{an} = 10$ ns, in the case of the bath coupled to the X degree of freedom. The locations of the first ($s_x$) and second ($s_+$) level crossings are shown, as well as the indices of the states.}
\end{figure}

To this end, we first introduce a bosonic Ohmic bath that is commonly used for dynamics simulations in quantum annealing and we solve the AME\cite{Albash2012, Albash2015cc, Albash2015a} numerically. We assume both qubits are coupled equally to the same independent baths, for which we specify the spectral density as
\begin{equation}
    \gamma(\omega) = 2 \pi \eta g^2 \frac{\omega \exp\left(-|\omega|/\omega_c\right)}{1 - \exp\left(-\beta \omega\right)}
    \label{eqn:bath-s-density}
\end{equation}
where $\beta=1/k_BT$ is the inverse temperature, $\omega_c$ is the cutoff frequency and $\eta g^2$ is the dimensionless coupling strength. In this work we restrict ourselves to values of these parameters similar to those used in theoretical studies of the DWave machine\cite{Albash2015a, Albash2015b, Mishra2018}, and refer to Refs. \cite{Chen2020} and \cite{Albash2015a} for the full details of the AME, since we only solve it numerically here. For the bath parameters we use $T=16$ mK, $\omega_c = 4 \times 2\pi$ GHz and $\eta g^2 = 10^{-4}$. The time-dependent Lindblad operators in the AME are defined as
\begin{equation}
    \hat{L}_{\alpha, \omega_{ba}} (t) = \langle E_b (t) | \hat{A}_\alpha | E_a (t) \rangle | E_a (t) \rangle \langle E_b (t) |
    \label{eqn:lindblad-ops}
\end{equation}
where $\hat{A}_\alpha$ is an operator of the system Hamiltonian, in this case Eq. (\ref{eqn:H-2-qubit-cluster}), and where $\omega_{ba}=E_b(t) - E_a(t)$ is the energy gap between instantaneous energy levels $a$ and $b$ at time $t$. An important feature of these operators is that they describe stochastic transitions in the energy eigenbasis of the system. This occurs in the weak coupling limit. In our simulations we use the operators $\hat{A}_{1,2} = \hat{\sigma}^x_{1,2}$ to describe relaxation, as their effect is to flip the qubit Z expectation. Separately we use $\hat{A}_{1,2} = \hat{\sigma}^z_{1,2}$ to describe dephasing, as their effect is to change the phase of the qubit in the XY plane. Due to our choice of energy scale $R=1$ GHz, we might expect all 16 Lindblad operators associated with $\hat{A}_{1,2}$ to play a role, owing to the fact that all transitions $\omega_{ba}$ have an energy in the region of $\omega_c$. However this isn't necessarily the case.

We first compare the probabilities of obtaining the ground state, for the given system-bath interaction, over a range of annealing times and with an initial state $\hat{\rho}(0) = |E_0(0)\rangle \langle E_0(0)|$. Figure \ref{fig:aqa-xz}(a) shows the ground state probability obtained in the AQA case, including the closed-system result for reference. We see that both relaxation (labeled X) and dephasing (labeled Z) play a significant role at long time scales, reducing the ground state probability before the dynamics are suitably adiabatic due to depopulation of the ground state. Figure \ref{fig:aqa-xz}(b) shows the overlap of the density matrix with the instantaneous eigenstates of the system and the location of the minimum gap $s_*$. As we would expect, the system begins to noticeably thermalize near $s=s_*$, as transitions are most likely there.

In the DQA case, the ground state probabilities are shown in Fig. \ref{fig:dqa-xz}(a), where the inset shows a closeup of the effect of a purely dephasing bath. We see here a stark contrast with the AQA case, where significant relaxation (X curve) occurs at a much smaller anneal duration, and where at longer durations $t_{an}>40$ ns repopulation of the ground state occurs, increasing the ground state probability again. In the case of dephasing (Z curve), the effect is surprisingly less significant compared to the AQA case. However, intuitively, the pronounced effect of relaxation in the DQA protocol is not unexpected, as the energy gap between the ground and first excited state is on average much smaller than in the AQA case, and thus we would expect a higher rate of depopulation/repopulation of the ground state. Indeed, this can be seen in Fig. \ref{fig:dqa-xz}(b), where depopulation of the ground state into the first excited state occurs at $t=0$. However, in the case of dephasing it is less obvious why there is an apparent insensitivity.

To better understand this, we consider the structure of the instantaneous eigenstates for the DQA protocol involving two qubits and analyze their effect on the Lindblad operators. We will restrict our analysis to consider only transitions between the ground and first excited states in the case $c_x=0$, justified by the fact that depopulation into the second excited state ($i=2$) is negligible as evidenced by Fig. \ref{fig:dqa-xz}(b). Under these conditions, the two lowest eigenstates of the Hamiltonian can be written in the form
\begin{subequations}
\begin{equation}
    |E_0(s)\rangle = |q_0(s)\rangle \otimes |\downarrow\,\rangle
    \label{eqn:dqa-state-0}
\end{equation}
and
\begin{equation}
    |E_1(s)\rangle = |q_0(s)\rangle \otimes |\uparrow\,\rangle
    \label{eqn:dqa-state-1}
\end{equation}
\end{subequations}
respectively, where $|q_0(s)\rangle$ denotes the lowest energy eigenstate of qubit 1. Qubit 2 will instantaneously flip twice in the ground state, owing to the absence of a transverse local field, once at $s_x$ and once at $s_+$. This means that, given our notation above, for $s < s_x$ and $s > s_+$, $|E_0(s)\rangle$ is the ground state, and for $s_x < s < s_+$, $|E_1(s)\rangle$ is the ground state. Using this notation, Eq. (\ref{eqn:lindblad-ops}) can be written as either
\begin{subequations}
\begin{equation}
    \hat{L}_{\alpha, \omega_{ba}} (t) = \langle q_0 (t) | \hat{A}_\alpha | q_0 (t) \rangle \langle b | \hat{I} | a \rangle | q_0 (t)a \rangle \langle q_0 (t)b|
    \label{eqn:lindblad-ops-q1}
\end{equation}
in the case that $\hat{A}_\alpha$ is an operator of qubit 1, and
\begin{equation}
    \hat{L}_{\alpha, \omega_{ba}} (t) = \langle q_0 (t) | \hat{I} | q_0 (t) \rangle \langle b | \hat{A}_\alpha | a \rangle | q_0 (t)a \rangle \langle q_0 (t)b |
    \label{eqn:lindblad-ops-q2}
\end{equation}
\end{subequations}
in the case that $\hat{A}_\alpha$ is an operator of qubit 2. Here $| q_0 (t)a \rangle \langle q_0 (t)b | = | q_0 (t) \rangle \langle q_0 (t) | \otimes |a \rangle \langle b |$, $\hat{I}$ is the $2\times 2$ identity matrix and $a$ and $b$ now distinguish the states based on the Z spin of qubit 2, i.e. $a,b \in \{\downarrow, \uparrow\}$. We can see immediately that in the first case, Eq. (\ref{eqn:lindblad-ops-q1}) is zero if $a \neq b$ due to the orthogonality of $|\downarrow\,\rangle$ and $|\uparrow\,\rangle$. This already rules out the possibility that any transition can be driven by a bath coupled to any degree of freedom belonging to qubit 1. In the case $a = b$, the associated Lindblad operator has a dephasing effect entirely determined by the expectation value of the chosen operator $\langle q_0 (t) | \hat{A}_\alpha | q_0 (t) \rangle$ and the amplitude of the bath correlation function at zero frequency $\gamma(0)$. In the second case,  Eq. (\ref{eqn:lindblad-ops-q2}), we can immediately see that only the choice $\hat{A}_\alpha = \hat{\sigma}^x_2$ yields non-zero terms for $a \neq b$, since only the off-diagonal elements of $\hat{\sigma}^x_2$ are non-zero. From this it is clear why relaxation plays an important role: the amplitude of the term $\langle b | \hat{\sigma}^x_2 | a \rangle$ is maximal for $a \neq b$. The case $a=b$ plays no role here as qubit 2 is always in a computational basis state under our assumptions.

Rather interestingly, this conclusion in the case of relaxation can be readily generalized if we consider $| q_0 (t) \rangle$ to describe many other qubits in their ground state, i.e. it doesn't matter what form this eigenstate takes as $\langle q_0 (t)| \hat{I}| q_0 (t) \rangle\langle b | \hat{\sigma}^x_2 | a \rangle=1$, and the result only depends on the bath spectral density at the transition frequency. Furthermore we may also conclude that all raising and lowering operators that result from coupling the bath to all other qubits are canceled in the two-level energy subspace, due to the $\langle b|\hat{I}|a\rangle$ term in Eq. (\ref{eqn:lindblad-ops-q1}), implying that as long as only transitions between the two lowest energy eigenstates are significant, and the AME applies, the relaxation rate of the entire system is completely determined by the target qubit.
\begin{figure}[t]
\includegraphics[scale=0.45]{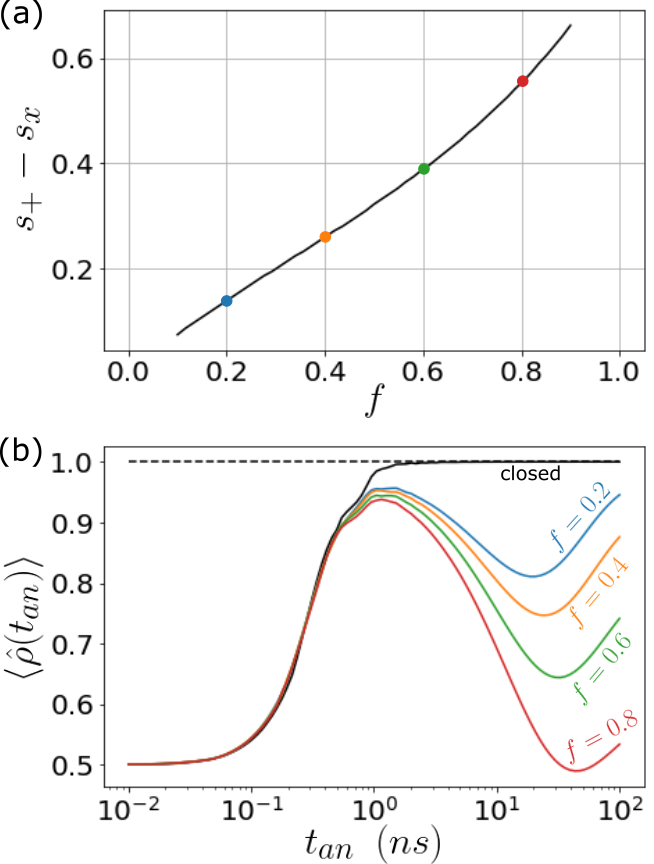}
\caption{\label{fig:dqa-x-f} The time interval $t_{an}(s_+ - s_x)$ the system spends in the first excited state is shown in (a), as a function of the frustration parameter $f$. Solutions of the ground state expectation of the density matrix solved in the AME as a function of total annealing time $t_{an}$ are shown in (b), for the selected values of $f$ represented as colored dots in (a).}
\end{figure}

To see why dephasing is strongly suppressed in this example, we must look at the behavior of the term $\langle q_0 (t) | \hat{A}_\alpha | q_0 (t) \rangle$. The case of no dephasing corresponds to the case where $| q_0 (t) \rangle$ is an eigenstate of $\hat{\sigma}^x$, since in this case, $\langle \pm | \hat{\sigma}^z | \pm \rangle = 0$, where $|\pm \rangle = \frac{1}{\sqrt{2}}(|0\rangle \pm |1\rangle)$. In fact, in our example, it can be verified numerically that for most of the evolution, $| q_0 (t) \rangle$ is almost an eigenstate of $\hat{\sigma}^x$, until near the end of the anneal. In other words, qubit 1 is strongly polarized along the X magnetization axis until near the end of the anneal, which explains the result seen in Fig. \ref{fig:dqa-xz}(a) (see Appendix \ref{app:qubit-x-mag}). Unlike the relaxation result however, this result is not readily generalized, as the structure of $| q_0 (t) \rangle$ is generally non-trivial. Therefore we would not expect this dephasing insensitivity to apply to all problems.

Another important property of the LSTF-DQA protocol when applied in an environment that drives relaxation is that the impact of relaxation can be mitigated if the time interval spent in the first excited state is reduced. Figure \ref{fig:dqa-x-f}(a) shows that decreasing the frustration $f$ in the system decreases the time interval $t_{an}(s_+-s_x)$ for fixed $s_x$, as the point $s_+$ occurs earlier in the anneal. Figure \ref{fig:dqa-x-f}(b) shows the overlap of the problem ground state with the density matrix at the end of the anneal, obtained by solving the AME as a function of $t_{an}$. We see that the effect of decreasing frustration is to reduce the impact of relaxation. Reducing the frustration has the effect of both reducing the time interval $t_{an}(s_+-s_x)$, but also reduces the average energy gap between the ground and first excited states. The latter property has a less pronounced impact on relaxation according to our results, and thus we do not discuss it further here. Some data is however presented in Appendix \ref{app:R-dependence}. It is worth noting that changing the energy scale could be beneficial for investigating resonance effects with the bath\cite{Marvian2015} experimentally. A particularly interesting application of the LSTF-DQA protocol is to perform multi-qubit interferometry experiments\cite{Munoz-Bauza2019,Karanikolas2020}, which could serve as a very sensitive probe of coherence in a quantum annealing processor. By adjusting the values of $c_x$ and $c_1$ in Eq. (\ref{eqn:X-sched-k}), it is possible to tune the size of the energy minima at $s_x$ and $s_+$ to induce oscillations in the ground state probability. This is also an application which we will focus on in future work.


%


\section{Conclusion}

We have shown that for specific types of optimization problems, i.e. those with frustration caused by inhomogeneous local Z fields, a DQA protocol can be formulated that exploits the suppression of quantum tunneling of frustrated spins to create arbitrarily small energy gaps in the annealing energy spectrum. This we call the locally-suppressed transverse-field (LSTF) DQA protocol. We first demonstrated, using a semiclassical approximation, how a frustrated qubit leads to the formation of a double-well potential when its transverse field is reduced, leading to a delocalized state at some point in $s$. In particular we showed that an arbitrarily small avoided crossing can be created, and that the gap can be closed entirely in the limit of a vanishing transverse field. We then presented a sketch for an $O(N)$ heuristic algorithm that could be used to find lower energy eigenstates of black-box problems under certain conditions. The performance of this algorithm was explored using a series of randomly generated instances, which showed that significant advantages can be obtained, particularly for problems in which phase transitions exist. We have also shown that the LSTF-DQA method is expected to be more sensitive to thermal relaxation than the AQA method for a given problem, through an analysis of the structure of the DQA eigenstates, and their role in forming the Lindblad operators of the AME. A single qubit dominates this effect even for larger systems under certain conditions, notably that only transitions between the ground and first excited states are important. The effect of dephasing in our DQA protocol was shown to be minimal for our two-qubit example. However no simple generalization to greater numbers of qubits was found. We therefore believe that this protocol will be of more use to determining the coherent evolution performance of near-term quantum annealing hardware, than for optimization applications. Despite this, we expect that there is some merit to applying this technique to both classical and quantum optimization in the near-term, particularly because it is inherently simplistic, has $O(N)$ performance, and can be used on existing novel quantum annealing hardware and in classical algorithms that simulate quantum annealing. Furthermore we believe it can be used to assess the structure of a black-box problem, for example testing for the presence of small energy minima.

\begin{acknowledgments}
We gratefully acknowledge Vicky Choi, Gioele Consani, Huo Chen, Robert Banks, Tameem Albash, Daniel Lidar and Evgeny Mozunov for useful comments and insightful discussion. The research is based upon work (partially) supported by the Office of the Director of National Intelligence (ODNI), Intelligence Advanced Research Projects Activity (IARPA), and the Defense Advanced Research Projects Agency (DARPA), via the U.S. Army Research Office Contract No. W911NF-17-C-0050. The views and conclusions contained herein are those of the authors and should not be interpreted as necessarily representing the official policies or endorsements, either expressed or implied, of the ODNI, IARPA, DARPA, or the U.S. Government. The U.S. Government is authorized to reproduce and distribute reprints for Governmental purposes notwithstanding any copyright annotation thereon.
\end{acknowledgments}

\bibliography{refs}

\appendix
\begin{figure}[t]
\includegraphics[scale=0.45]{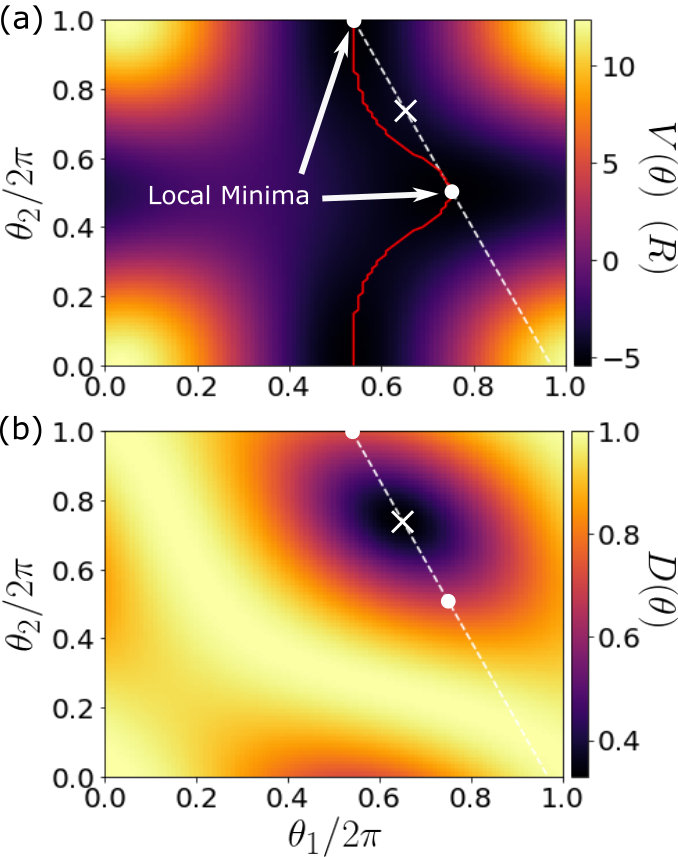}
\caption{\label{fig:potential-3d} The semiclassical potential Eq. (\ref{eqn:V-2-qubit-cluster}) solved on a grid of $\theta_1$ and $\theta_2$ values and at $s=s_+$ is shown in (a). The tracenorm distance Eq. (\ref{eqn:D-2-qubit-cluster}) solved for the same parameters is shown in (b). The position of the local minima of Eq. (\ref{eqn:V-2-qubit-cluster}) are indicated in (a) and denoted as white dots in both figures. The position of the global minimum of Eq. (\ref{eqn:D-2-qubit-cluster}) is denoted as a white cross in both figures. The line $\theta$, Eq. (\ref{eqn:potential-line}), is shown as the white dotted line in both plots, along which lie the two local minima of the semiclassical potential and the global minimum of the tracenorm distance. The red line in (a) indicates the locus of the minimum potential energy with respect to changes in $\theta_2$.}
\end{figure}

\section{Semiclassical Potential\label{app:semiclassical}}

We describe here how the results shown in Fig. \ref{fig:potential}(a) and (b) were obtained. Consider Figs. \ref{fig:potential-3d}(a) and \ref{fig:potential-3d}(b) which show the values obtained for Eqs. (\ref{eqn:V-2-qubit-cluster}) and (\ref{eqn:D-2-qubit-cluster}) respectively, as a function of $\theta_1$ and $\theta_2$ at $s=s_+$. We are interested in the local (including the global) minima of the semiclassical potential. To find these local minima, we first compute the locus of the minimum potential energy with respect to changes in $\theta_2$, indicated by the red line in Fig. \ref{fig:potential-3d}(a). We then take the first numerical derivative of the energy with respect to $\theta_2$ along the locus points. The location of the zero crossings at which this derivative is increasing correspond to the location of the local minima, shown as the white dots in the figure. We repeat this process for different values of $s$ to trace the energy of the local minima as a function of $s$, which gives us the data required for Fig. \ref{fig:potential}(a). To obtain the data shown in Fig. \ref{fig:potential}(b), we define the straight line $\theta(s, \theta_1, \theta_2)$ that intersects the location of both local minima. We denote the coordinates of the first local minimum as $\theta_{1,a}$, $\theta_{2,a}$ and at the second minimum as $\theta_{1,b}$, $\theta_{2,b}$. We can express the line $\theta$ as a function of either $\theta_1$ or $\theta_2$. Therefore we can express the line as $\theta(s, \theta_1, \theta_2) \equiv \theta_2'(s, \theta_1)$ where
\begin{equation}
    \theta_2'(s, \theta_1) = \frac{\theta_{2,a}(s) - \theta_{2,b}(s)}{\theta_{1,a}(s) - \theta_{1,b}(s)} \left(\theta_1 - \theta_{1,a}(s)\right) + \theta_{2,a}(s)
    \label{eqn:potential-line}
\end{equation}
where we have made explicit the dependence on $s$, and $\theta_2'$ is the value of $\theta_2$ when $\theta_1$ is on the line $\theta$. We then solve Eq. (\ref{eqn:V-2-qubit-cluster}) along $\theta$ and express it as a function of $\theta_2'$ to obtain the semiclassical tunneling potential seen as the green line in Fig. \ref{fig:potential}(b). We also solve Eq. (\ref{eqn:D-2-qubit-cluster}) along $\theta$ to obtain the tracenorm distance seen as the black dotted line in Fig. \ref{fig:potential}(b). Notice that the global minimum of the tracenorm distance, shown as the white cross in Figs. \ref{fig:potential-3d}(a) and \ref{fig:potential-3d}(b), lies on the line $\theta$.
\begin{figure}[t]
\includegraphics[scale=0.5]{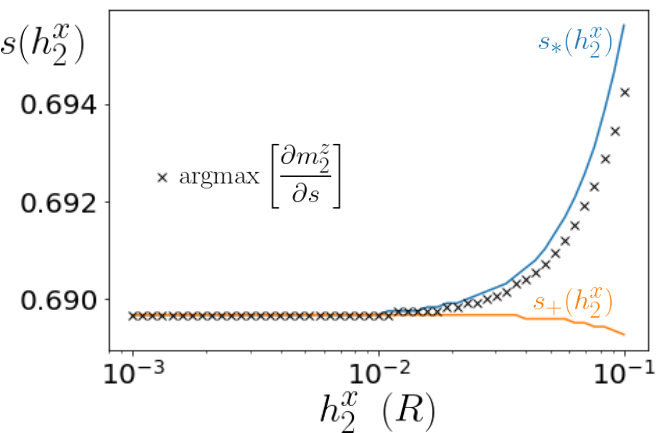}
\caption{\label{fig:sstarsplus} The positions of the minimum gap at $s=s_*$ and the zero crossing of the Z magnetization of qubit 2, $s=s_+$, as a function of the suppressed transverse field $h^x_2$. The black crosses denote the values of $s$ at which the rate of change of the qubit Z magnetization $\partial m^z_2 / \partial s$ is greatest, which correlate well with the position of the minimum gap. The parameter values used were $f=0.8$ and $J^{zz} = h^z_1 = R$.}
\end{figure}

\section{Divergence of $s_*$ and $s_+$\label{app:sstarsplus}}

One might expect $s_*$, the location at which the minimum gap occurs, to be equal to $s_+$, the location at which the ground-state Z-magnetization crosses zero. We show here that this isn't necessarily the case. Figure \ref{fig:sstarsplus} shows the values of $s_*$ and $s_+$ as a function of $h^x_2$ for the two-qubit Hamiltonian described by Eq. (\ref{eqn:H-2-qubit-cluster}). As $h^x_2$ becomes larger, the difference between $s_*$ and $s_+$ grows. This behavior was also observed in recent work by Albash and Kowalsky \cite{Albash2021}. In Fig. 11 of their work, the points which we define as $s_*$ and $s_+$ are shown to converge in the $p$-spin model, as $p$ is increased. Similarly the minimum gap occurs after the zero-crossing of the magnetization in their results. It is also noted in their work that the location of the minimum gap corresponds to the point at which the slope of the Z magnetization is greatest. We verified this numerically with our 2-qubit example, represented by the black crosses in Fig. \ref{fig:sstarsplus}, where indeed a good correlation can be observed between the slope of $m^z_2$ and the location of the minimum gap $s_*$.

\section{7-Qubit Random Problem Instances}

\subsection{Graph Generation Algorithm\label{app:random-instance-graphs}}

To generate a random connected graph, we first start with a fully connected 7-vertex graph. We then iteratively choose random edges to remove one-by-one. For each edge to remove, we first check if the graph is connected. If it is not, we choose a different edge and restart. If the graph is still connected, we proceed with the choice and choose a new edge to remove, until the desired number of edges remains. To ensure the generated graphs are non-isomorphic with respect to one another, we compare the newly generated graph with all those in a list. If one is found to be isomorphic, the newly generated graph is discarded and the process repeats until we have obtained 10 non-isomorphic graphs. We used the \texttt{python} package \texttt{networkx}\cite{Hagberg2008} to create and manipulate graphs, including checking for isomorphism and connectivity.
\begin{figure}[t]
\includegraphics[scale=0.3]{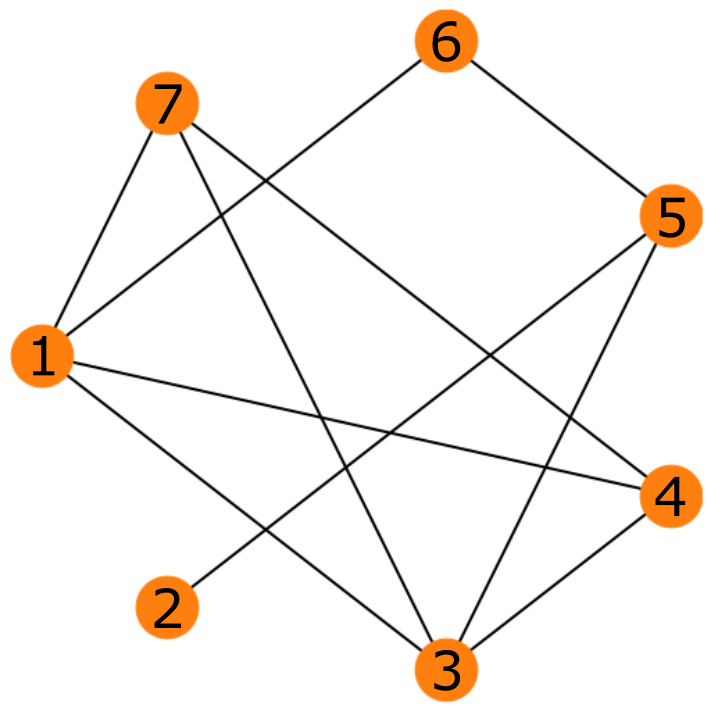}
\caption{\label{fig:random-G} The 10-edge graph used for the random instance analysis, discussed in section III B of the main text.}
\end{figure}

\subsection{Instance Parameters\label{app:random-instance-vals}}

The graph that was used for the random problem example is shown in Fig. \ref{fig:random-G}. This is one of ten graphs generated by the algorithm described in Appendix \ref{app:random-instance-graphs}. The values of $h^z_i$ used for the random problem example are
\begin{eqnarray*}
h^z_1 = 1\,\,\mathrm{GHz}, \\ h^z_2 = -0.32610452\,\,\mathrm{GHz}, \\ h^z_3 = 0.16998698\,\,\mathrm{GHz}, \\ h^z_4 = -0.12109217\,\,\mathrm{GHz}, \\  h^z_5 = -0.58725647\,\,\mathrm{GHz}, \\ h^z_6 = 0.19980255\,\,\mathrm{GHz}, \\ h^z_7 = -0.4370849\,\,\mathrm{GHz}.
\end{eqnarray*}
which were drawn from identical Gaussian distributions with a variance of 1 and then normalized such that the largest field has a magnitude $R$, which in this case is set at $R = 1$ GHz. The values of $J^{zz}_{ij}$ were all set to $-0.5 R$ and all the transverse fields $h^x_i$ were set to $R$, as described in the main text.
\begin{figure}[t]
\includegraphics[scale=0.45]{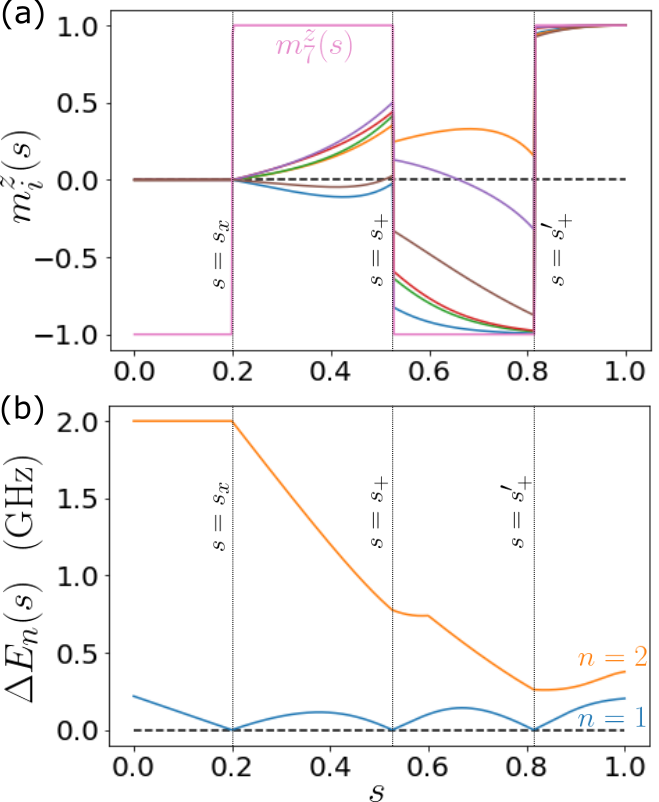}
\caption{\label{fig:random-inst-double} In (a) is shown the magnetization expectation values of each qubit when using LSTF-DQA applied to qubit 7 ($k=7$) of the graph shown in Fig. \ref{fig:random-G}. In (b) the resulting energy spectrum is shown. In both figures the points $s=s_x$, $s=s_+$ and $s=s_+'$ are indicated.}
\end{figure}

\subsection{Observed Features\label{app:random-instance-props}}

We discuss here some additional observed features of the random instance described in the main text, focusing on the data shown in Fig. \ref{fig:random-magz}(a). In this figure, we can clearly see that a few qubit Z magnetization values undergo a double zero-crossing during the evolution for the AQA process (the pink and purple lines). We show in Figs. \ref{fig:random-inst-double}(a) and (b) that application of LSTF-DQA to such a qubit can result in a triple flip of the qubit Z magnetization, and consequently a triple crossing of the ground and first excited states. These crossing points are indicated at $s=s_x$, the second at $s=s_+$ and a third at a point we denote $s=s_+'$. This suggests it is not always necessary to create an additional crossing at $s=s_x$, and also suggests that the way in which local minima in the potential compete can be highly non-trivial. Notice also from Fig. \ref{fig:random-inst-double}(a) that the other qubits must clearly participate in the tunneling in some way, since they are all influenced by the flip events of the selected qubit ($k=7$ here). This also occurs to a lesser extent when qubit 1 is the target qubit, as seen in Fig. \ref{fig:random-magz}(b), where the pink line appears to no longer cross zero near $s=0.4$ following the application of LSTF-DQA.

We can see this also in the context of the 2-qubit example when the transverse field of qubit 2 is suppressed. Considering Fig. \ref{fig:potential-3d}(a), we see that a sudden rotation of qubit 1 from the X to the Z axis accompanies the full Z expectation flip of qubit 2. These correlated events are not at all obvious unless the transverse field of the target qubit is suppressed. It is striking that the suppression of a single transverse field can reveal these otherwise hidden correlations. Finally, note that from a computational perspective, we expect that omitting the crossing at $s=s_x$ and simply suppressing a selected transverse field will not be very effective, since such a double crossing of the Z magnetization in the AQA spectrum is expected to be a rare occurrence, and furthermore it does not consistently yield the described behavior. We might expect to observe the same features revealed in Figs. \ref{fig:random-inst-double}(a) and (b) when selecting qubit 5 ($k=5$), for which the Z magnetization corresponds to the purple curve in Figs. \ref{fig:random-magz}(a), \ref{fig:random-magz}(b) and \ref{fig:random-inst-double}(a). However this is not the case despite the double zero-crossing of the Z magnetization. In that case the Z-magnetization no longer changes after $s=s_x$, and we observe a `softening' of the phase transition, whereby the minimum gap is increased in size.
\begin{figure}[b]
\includegraphics[scale=0.5]{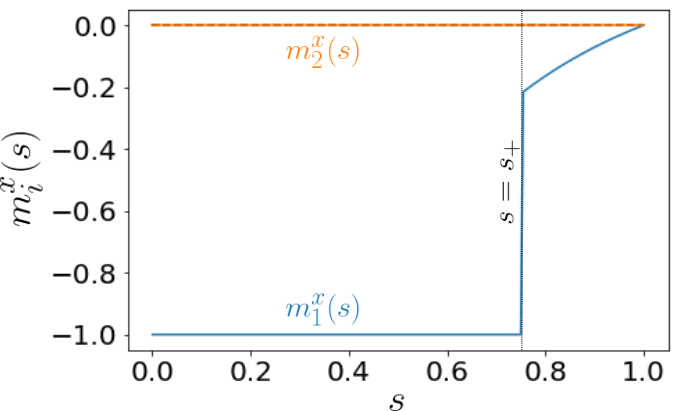}
\caption{\label{fig:x-mag} The magnetization of each qubit in the frustrated two-qubit problem, when $f=0.8$ and $c_x=0$. For $s>s_+$, qubit 1 is no longer fully X polarized.}
\end{figure}

\section{Qubit X Magnetization\label{app:qubit-x-mag}}

We argued that dephasing in our two-qubit example (Eq. (\ref{eqn:H-2-qubit-cluster})) was strongly suppressed due to the fact that $| q_0 (t) \rangle$, associated with qubit 1, is very closely an eigenstate of the $\hat{\sigma}^x$ operator throughout the duration of the anneal. To see this, we show the ground state magnetization of both qubits $m^x_{1,2} = \langle E_0 | \hat{\sigma}^x_{1,2} | E_0 \rangle$ in Fig. \ref{fig:x-mag}. From this plot the ground-state X-magnetization of qubit 1, encoded in the state $| q_0 (t) \rangle$, is maximal up to the point $s=s_+$. For $s>s_+$ there should be a finite dephasing effect due to qubit 1. The ground-state X-magnetization of qubit 2 is zero throughout the anneal, as expected in this case since $c_x = 0$. We emphasize here and in the main text that the strong X-magnetization of qubit 1 throughout most of the anneal is a feature of this simple problem. The reason this occurs is firstly that it is energetically favorable for qubit 1 to rotate to the X-axis early in the anneal where the transverse field dominates, and secondly this allows the energy penalty caused by the choice of $J^{zz}$ to be removed between $s=0$ and $s=s_+$. Beyond the point $s=s_+$, qubit 2 is flipped back into the favorable direction, which makes the magnetization of qubit 1 strongly Z-polarized to this configuration being more energetically favorable than if it were X-polarized.

\section{Energy Dependence of Relaxation Effects\label{app:R-dependence}}
\begin{figure}[b]
\includegraphics[scale=0.5]{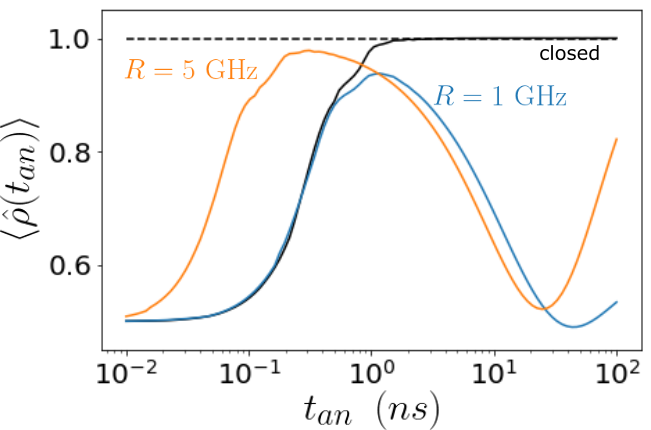}
\caption{\label{fig:x-R-relax} The probability of obtaining the ground state at the end of an anneal for the DQA protocol, calculated by solving the AME for the Ohmic bath coupled to the X degrees of freedom for different energy scales $R=1$ GHz and $R=5$ GHz. The closed system result is shown as the solid black line for reference.}
\end{figure}

Figure \ref{fig:x-R-relax} shows the probability of obtaining the ground state at the end of an anneal using LSTF-DQA for two different values of the energy scale $R$ in the Hamiltonian of Eq. (1). Only the X degrees of freedom are coupled to the Ohmic bath. The effect of increasing the energy scale does not have a great impact on the relaxation and depopulation rates. Rather, by increasing the energy scale we have decreased the anneal duration $t_{an}$ required for adiabatic dynamics. At the shorter timescales associated with $R=5$ GHz, the system has very little time to thermalize, and thus we obtain a slightly larger peak in ground state probability than for $R=1$ GHz. The depopulation of the first excited state also appears to occur over a shorter timescale in this case.

\end{document}